\documentclass[runningheads]{llncs}

\authorrunning{Yusof et al.}

\usepackage{acronym}
\usepackage{xspace}
\usepackage{svg}
\usepackage{graphicx}
\usepackage{booktabs}
\usepackage{multirow}
\usepackage{listings}
\usepackage{framed}
\usepackage{pifont}
\usepackage{url}
\usepackage[T1]{fontenc}

\newcommand{\ie}{i.e.,\xspace}
\newcommand{\eg}{e.g.,\xspace}

\newcommand{\sone}{AS1\xspace}
\newcommand{\stwo}{AS2\xspace}

\acrodef{apt}[APT]{Advanced Persistent Threats}
\acrodef{ics}[ICS]{Industrial Control Systems}
\acrodef{plc}[PLC]{Programmable Logic Controller}
\acrodef{scada}[SCADA]{Supervisory Control and Data Acquisition}
\acrodef{it}[IT]{Information Technology}
\acrodef{ot}[OT]{Operational Technology}
\acrodef{dfir}[DFIR]{Digital Forensics and Incident Response}
\acrodef{ioc}[IoC]{Indicator of Compromise}
\acrodef{ccdcoe}[CCDCOE]{Cooperative Cyber Defence Centre of Excellence}
\acrodef{dos}[DoS]{Denial of Service}
\acrodef{ddos}[DDoS]{Distributed Denial of Service}
\acrodef{fdi}[FDI]{False Data Injection}
\acrodef{fci}[FCI]{False Control Injection}
\acrodef{mitm}[MITM]{Man-in-the-Middle}
\acrodef{hmi}[HMI]{Human-Machine Interface}
\acrodef{rtu}[RTU]{Remote Terminal Unit}
\acrodef{fdc}[FDC]{Field Device Controller}
\acrodef{ddos}[DDoS]{Distributed Denial of Service}
\acrodef{ttp}[TTPs]{Tactics, Techniques and Procedures}
\acrodef{ioc}[IoC]{Indicator of Compromise}
\acrodef{cti}[CTI]{Cyber Threat Intelligence}
\acrodef{cue}[CUE]{Cluster User Emulation System}
\acrodef{c2}[C2]{Command and Control}
\acrodef{ssl}[SSL]{Secure Sockets Layer}
\acrodef{cps}[CPS]{Cyber-Physical Systems}

\title{Signals and Symptoms: ICS Attack Dataset From Railway Cyber Range}

\author{
    Anis Yusof\inst{1} \and
    Yuancheng Liu\inst{2} \and
    Niklaus Kang\inst{2} \and
    Choon Meng Seah\inst{2} \and
    Zhenkai Liang\inst{1} \and
    Ee-Chien Chang \inst{1}
}

\institute{
    School of Computing, National University of Singapore \\
    \email{\{anis, liangzk, changec\}@comp.nus.edu.sg}
    \and
    National Cybersecurity R\&D Lab, National University of Singapore \\
    \email{\{yc\_liu, niklausk, seahcm\}@nus.edu.sg}
}

\begin{document}

\maketitle
\begin{abstract}
The prevalence of cyberattacks on \ac{ics} has highlighted the necessity for robust security measures 
and incident response to protect critical infrastructure. This is prominent when \ac{ot} 
systems undergo digital transformation by integrating with \ac{it} systems to enhance 
operational efficiency, adaptability, and safety. 
To support analysts in staying abreast of emerging attack patterns, there is a need for 
\ac{ics} datasets that reflect indicators representative of contemporary cyber threats.
To address this, we conduct two \ac{ics} cyberattack simulations to showcase the impact of 
trending \ac{ics} cyberattacks on a railway cyber range that resembles the railway infrastructure. 
The attack scenario is designed to blend trending attack trends with attack patterns observed 
from historical \ac{ics} incidents. The resulting evidence is collected as datasets, serving as 
an essential resource for cyberattack analysis. This captures key indicators that are relevant 
to the current threat landscape, augmenting the effectiveness of security systems and analysts 
to protect against \ac{ics} cyber threats. 
\end{abstract}

\section{Introduction}

Cyberattack often targets \ac{ics} and cause disruptions to industrial operations. 
This is especially crucial for critical infrastructure, whose
operations are vital to society. The impact of an \ac{ics} cyberattack could lead to 
life-threatening situations. One such critical infrastructure is the railway system
comprising of interconnected systems and devices across the \ac{it} and \ac{ot} segments. 
The railway systems has evolved to include complex \ac{ot} systems to meet the growing demand 
of mass transportation. While such advancement is targeted to enhance the safety and reliability
of railway infrastructure, complex interdependent systems impose a significant cybersecurity 
challenge in defending and detecting threats. Furthermore, the segmented
systems increase the complexity of understanding the malicious behaviors when a cyberattack
occurs in railway systems. This raises the need for security analysts to be well equipped so
as to promptly react and analyze a cyber incident, minimizing the operational impact without 
endangering human safety.

Security analysts generally rely on past incidents as a case study to conduct digital forensics 
and gain a better understanding about existing threats.
To improve the overall effectiveness of analysis, one method is to conduct a cyber exercise
by simulating various threats in the rail infrastructure~\cite{7847966,10.1145/3384217.3385623} 
and assessing their response based on playbook~\cite{10.1145/3600160.3600195}. Some examples 
of threats include ransomware and \ac{dos} attacks which disrupt the operational and safety 
aspect of the infrastructure~\cite{FERNANDES2025100305}. Cyber exercise provides a platform 
for analysts to familiarize themselves with the forensics tools and validate their 
skillsets. Additionally, any lapses in existing workflow can be detected and rectified, improving 
the overall communication and coordination across various teams (\eg management, operators, engineers).
However, conducting a cyber exercise is often laborious and costly, requiring significant efforts
to realize the simulated threats. Furthermore, the datasets resulting from such cyber exercises are often
used within their context and are not made available for future use.

Analyzing real-world cyberattack data extracted from an actual railway system would provide valuable 
insights. This includes the actual \ac{ioc} and attack patterns observed from the data. However, 
real-world data are generally unavailable due to security and privacy concerns. Furthermore, 
conducting a simulated cyberattack on real-world systems to obtain useful data is infeasible 
due to the associated risks. This includes the risk of operational downtime during a train malfunction 
and the cascading effect of failures due to other fragile components. To overcome these challenges, 
one method is to replicate the rail infrastructure as a digital twin. By modeling the railway systems 
accurately, the digital twin provides a safe and controlled virtual environment for investigation 
without jeopardizing real-world operations. The digital twin also provides better control and monitoring 
capabilities, allowing analysts to execute their workflow and assess the impact based on the data
extracted from the railway systems. However, developing a digital twin involves significant cost and 
complexity, further aggravating the challenge of making useful \ac{ics} datasets accessible.
Despite this, the lack of quality data for \ac{ics} cyberattacks limits the understanding of the 
threat landscape~\cite{10267968,9471765}, hampering the ability to effectively defend against 
trending attack patterns.

This raises the need to support cyberattack analysis with datasets that are an approximation of 
real-world attack scenarios which can then be used for various analysis, including 
tracing attacks, identifying vulnerabilities, and mitigating future risks. However, existing studies
has highlighted the need for the simulation testbed to be realistically designed so as to avoid having
highly dependent features and experimental artifacts in the generated data~\cite{10629000,liu25sp}. 
In this work, our goal is to generate \ac{ics} attack dataset that is useful for various security 
applications. Specifically, we design two realistic cyberattack scenarios based on attack patterns 
observed from actual historical \ac{ics} cyberattacks. These cyberattack scenarios are customized for 
a large-scale international cyber exercise organized by a European-based center of excellence specializing 
in international research, training, and capacity in building cybersecurity and cyber defense. 
The cyberattack is executed in a railway cyber range, generating relevant evidence which is then 
collected as a dataset. Additionally, we share the challenges faced in designing the cyberattack scenario 
and collecting the relevant evidence. To support future research efforts, we collect the 
evidence as datasets and share them in a public repository~\footnote{https://doi.org/10.5281/zenodo.15536351}.
To summarize, the contribution of this work is as follows:
\begin{itemize}
    \item We share the design of our railway cyber range that enables experimentation with cyberattacks on railway infrastructure.
    \item We present the design of two cyberattack scenarios that are customized for a large-scale international cyber exercise on \ac{ics}.
    \item We publicly release two \ac{ics} attack datasets resulting from the cyberattack scenarios to support future research in this field.
\end{itemize}

The remainder of this paper is structured as follows. In Section~\ref{sec:background}, we introduce the background of
\ac{ics} cyberattacks and the railway cyber range. In section~\ref{sec:methodology}, we introduce the 
methodology of our study in designing the cyberattack scenario and experimentation to extract the 
relevant evidence. In Section~\ref{sec:experiment}, we empirically evaluate the dataset collected from 
the cyberattack scenario and presented a case study of analyzing the dataset. In Section~\ref{sec:related-work}, 
we identify related works on \ac{ics} cyberattack simulation platforms and generating simulated datasets. 
Section~\ref{sec:conclusion} concludes the paper.

\section{Background}
\label{sec:background}

In this section, we provide the context of executing cyberattack simulation on a railway cyber range. 
We first introduce the threat landscape of \ac{ics}, then we introduce the railway infrastructure 
that simulates real-world trains and signalling systems.

\subsection{Threat Landscape}
Understanding the threat landscape of critical infrastructure is 
the key to designing attack scenarios that accurately represent current \ac{ics} cyberattacks. 
Historically, cyberattacks conducted by \ac{apt} are known to target industries that are important to 
national security and the economy. This includes energy facilities, transportation, telecommunications, 
and healthcare. Consequently, the cyberattacks cause disruptions and potentially allow them
to conduct cyber espionage. Despite critical infrastructure being well protected, \ac{apt} cyberattacks 
are targeted and highly motivated by their goal, which makes them challenging to detect as they 
stealthily conduct longitudinal multi-stage attacks on the critical infrastructure~\cite{app13063409}.

The \ac{ics} infrastructure is categorized into \ac{it} and \ac{ot} segments. Based on the operational needs,
systems in the \ac{ot} segment are designed to prioritize high availability and safe operations. Additionally, 
the systems in \ac{ics} are generally isolated and do not provide direct communication with the 
public~\cite{10.1007/s10207-024-00856-6}, thus neglecting the need for security. This effectively reduces 
the attack surface for critical infrastructure, as public-facing connections provide more opportunities 
for threat actors to establish a foothold. 

The systems in the \ac{it} segment may need to interact with the \ac{ot} systems for various functionalities 
(\eg remote monitoring and access). Furthermore, this is especially prominent where \ac{ics} infrastructure 
undergoes digital transformation to make it more intelligent, exposing these \ac{ot} systems to typical 
\ac{it}-related cyber threats. This effectively invalidates existing assumptions where air-gapped 
\ac{ot} systems are isolated and inherently secure. As a result, \ac{ics} systems are facing an emerging threat
originating from the \ac{it} segment. 

The initial 
attack vector is a vital entry point for \ac{apt} to gain an initial foothold into the critical infrastructure. 
Some of the patterns that are commonly observed for the initial access include supply chain compromise 
(\eg Dragonfly \ac{apt}) and spearphishing (\eg APT33, APT44). Further aggravating the situation, the connectivity from \ac{it} segment has enable existing 
\ac{it}-related threats (\eg LockBit ransomware campaign) to make their way into critical infrastructure. 
This suggest the shift of focus of threat actors from achieving financial gain through ransom to causing 
disruptions on \ac{ics} systems. This reflects the consequence of intersecting \ac{it} and \ac{ot} segments, 
increasing the complexity of detecting and preventing \ac{ics} cyberattacks. Concurrently, such complex 
characteristics of cyberattacks provide the basis for designing realistic attack scenarios.

\subsection{Railway Cyber Range}

\begin{figure}[htp!]
    \centering
	\includegraphics[width=\textwidth]{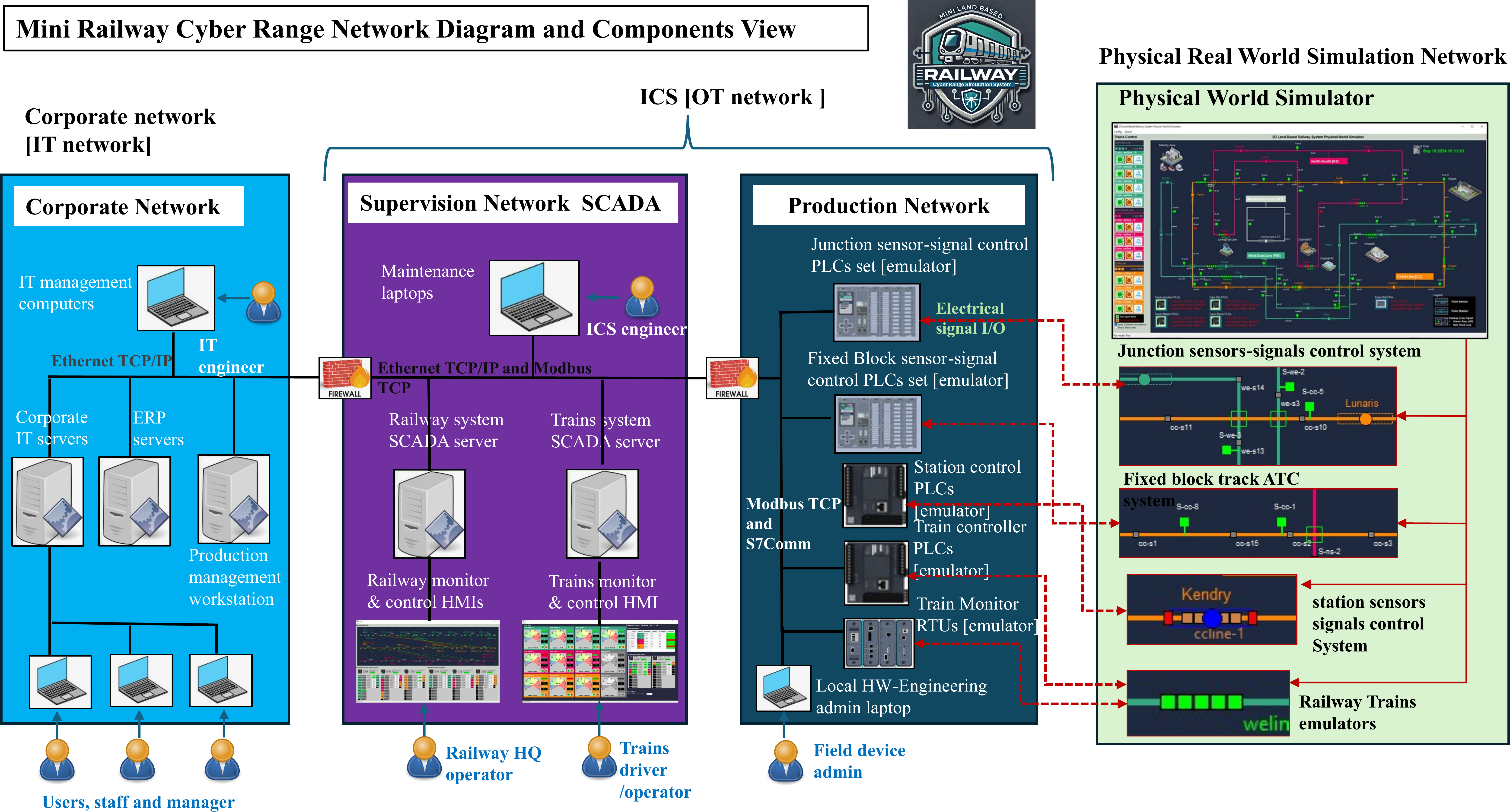}
    \caption{An overview of the railway cyber range}
    \label{fig:digital-twin-overview}
\end{figure}

Experimentations enable analysts to gain insights into various cyberattack scenarios 
across \ac{it} and \ac{ot} systems. To facilitate experiments on a railway infrastructure, we 
built a simulated railway \ac{it}/\ac{ot} system as a railway cyber range to conduct cybersecurity 
experimentation. The railway cyber range models the railway infrastructure and has emerged as a 
useful mechanism to simulate \ac{ics} cyberattack, effectively providing a safe environment for 
experimentation with cyber threats. As reflected in Figure~\ref{fig:digital-twin-overview}, 
the platform consists of four network systems, namely the corporate network, supervision SCADA network, 
production network, and physical real-world simulation network. The systems in both \ac{it} and \ac{ot}
segments serve distinct purposes. The \ac{it} segment consists of a corporate network that 
focuses on business processes (\eg workstations, web server, business applications).
The \ac{ot} segment consists of both supervision and production networks which control the equipment 
and operational processes (\eg sensors, actuators). The real-world network emulates the physical effects 
that the devices have caused in the real world. This platform is designed as a miniature
railway \ac{it}/\ac{ot} system that reflects the fundamental operational logic and is used 
as a cyber range to conduct cyber exercises.

\subsubsection{\ac{it} Systems --}
To replicate a realistic \ac{it} system in the railway platform, we establish three key
components that form the \ac{it} system. They consist of \ac{it} computing environment, 
the execution of realistic organizational activities, and conducting attack behaviors. 
The \ac{it} environment simulates a typical corporate network of a railway organization. 
This environment consists of various business-oriented computing systems such as workstations, 
firewalls, routers, and switches. Based on real-world computing environment, diverse day-to-day 
activities from the respective business units are conducted using the \ac{it} systems. 
To generate benign activities on the computing environment and the corresponding network traffic,
the activities of railway staff are virtually simulated. This includes actions 
made by a railway operator, train driver, safety officer, \ac{it} support engineer, and 
administrative officer. In addition to benign behaviors, we simulate the malicious activities to 
mimic attack scenarios throughout the \ac{it} systems. This includes the process of sending 
phishing emails, conducting \ac{fci}, \ac{fdi}, \ac{mitm}, and \ac{ddos} attacks.

\subsubsection{\ac{ot} Systems --}
The \ac{ot} system in the platform replicates the \ac{ics} components that control the railway
system. The \ac{ot} environment consists of two networks, namely the \ac{scada} network and 
the production network. The \ac{scada} network consists of control and monitoring systems.
This includes the \ac{scada} historian server that serves as a centralized repository for
data collected from \ac{scada} systems. Additionally, this network also contains \ac{hmi} for
monitoring and maintenance computers used by system operators and engineers, respectively.
The production network contains \ac{fdc} that simulates a realistic representation of
production environment in a railway system. This includes the simulation of \ac{plc} 
and \ac{rtu} devices.

The \ac{ics} components in the \ac{ot} systems are functionally categorized into two subsystems, 
namely the railway signaling
system and the train control system. The track signaling system simulates the control of track
junctions and railway stations. To automate the control of the tracks and junctions, the signaling
system reads the simulated electrical signals from the devices in the real-world network. 
For example, the devices in the real-world network indicate the train's position
on the track. This allows the signaling system to control the fixed block signaling, manage
passing trains through different track junctions, and guide trains to dock and depart from
railway stations.

\begin{figure}[htp!]
    \centering
	\includegraphics[width=\textwidth]{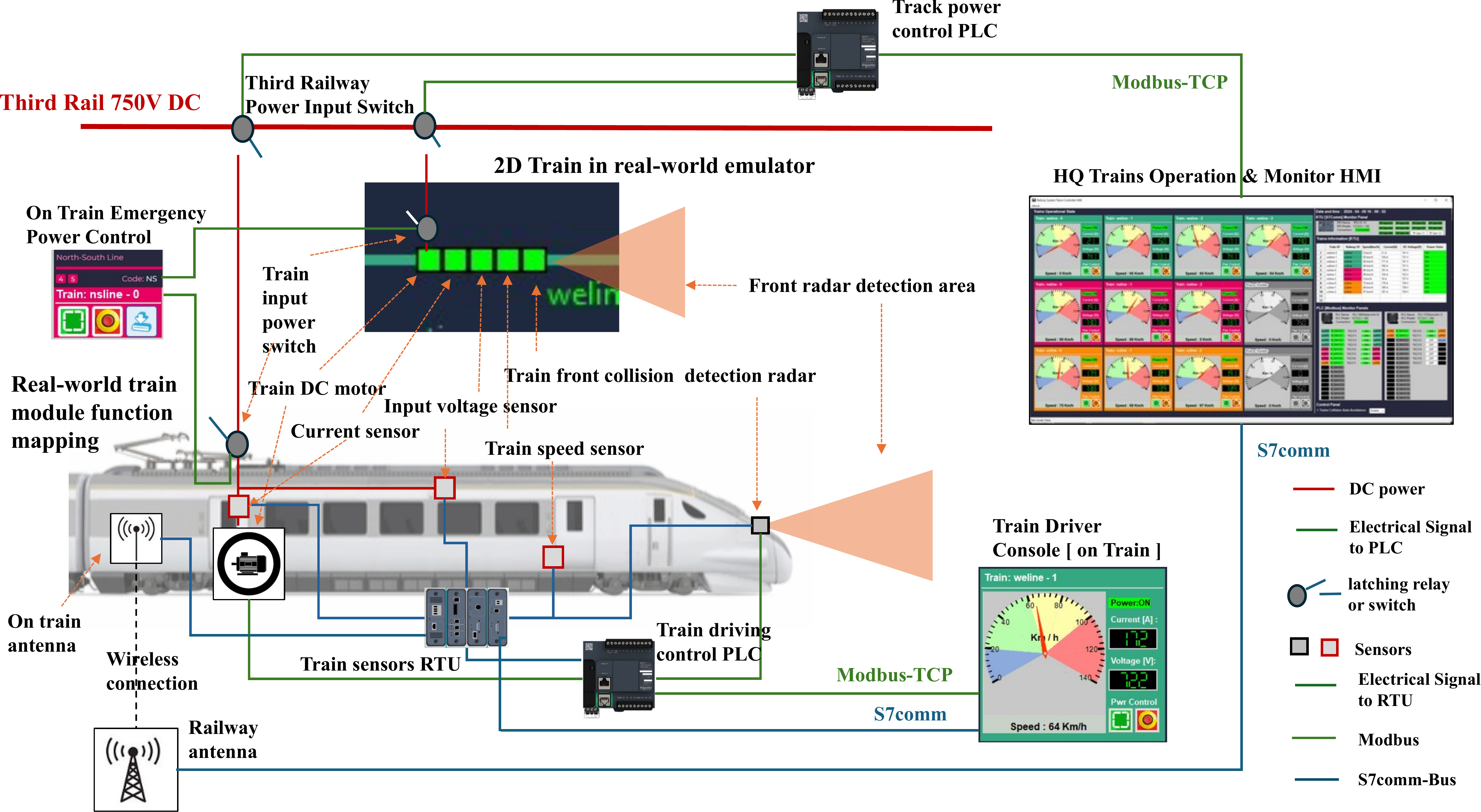}
    \caption{An overview of the train control system}
    \label{fig:train-control-system}
\end{figure}

As reflected in Figure~\ref{fig:train-control-system}, the train control system replicate the 
train management by simulating the \ac{plc} and \ac{rtu} devices. 
The \ac{plc} devices are used to manage the third rail, train signaling, and collision avoidance
mechanisms. Additionally, the \ac{rtu} devices are used to collect operational sensor data from
the train such as throttle and braking, current speed, input voltage, and motor current. This allows
the train control system to detect the state of track signals, thus making coordinated decisions
for the train to pass through junctions and make a stop at stations. Furthermore, the train control
system monitors the train ahead to avoid collision and adheres to the fixed block states of
railway tracks. The train control system also communicates with the railway headquarters and provides
monitoring and control capabilities on the train driver's console.

\subsubsection{Railway Real World Emulation --}

\begin{figure}[htp!]
    \centering
    \includegraphics[width=\textwidth]{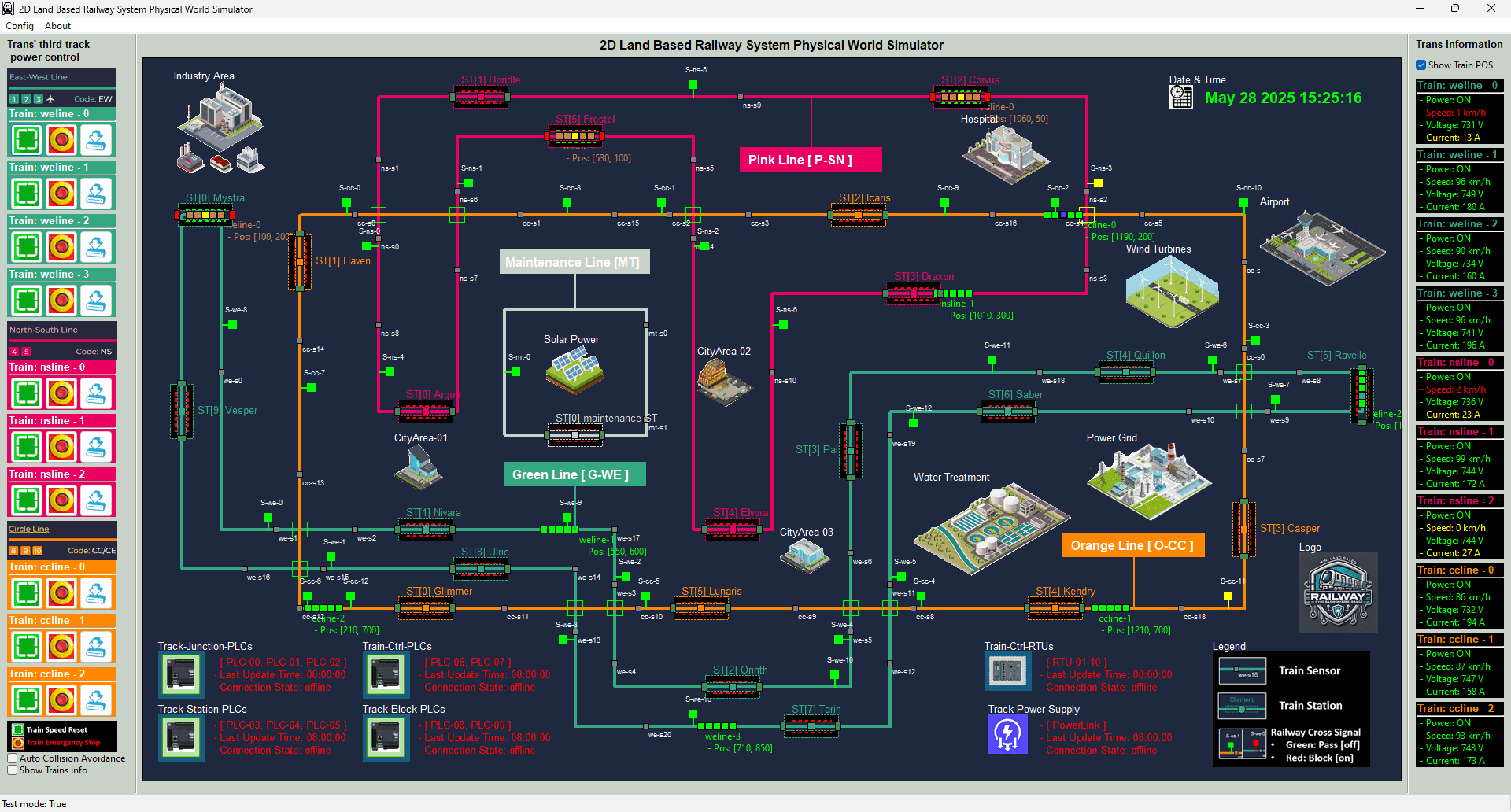}
    \caption{Visualization of the railway cyber range}
    \label{fig:train-visualization}
\end{figure}

The railway trains are simulated based on the interconnected systems across both \ac{it} and \ac{ot}
networks. To visualize the train network, we provide a graphical user interface for the users
of this platform to monitor and control the system as shown Figure~\ref{fig:train-visualization}. 
This includes the visualization of physical
real-world scenarios of trains traversing the tracks and docking at stations. The railway emulator
features ten trains across four tracks. This includes the simulation of the track signaling system which 
simulates track junction control and railway stations. By having the visualization, the emulator aims 
to deliver a realistic and dynamic railway environment that serves as a cyber range to perform 
experimentation on a railway platform.

\section{Methodology}
\label{sec:methodology}

\begin{figure*}[htp!]
    \centering  
	\includegraphics[width=\textwidth]{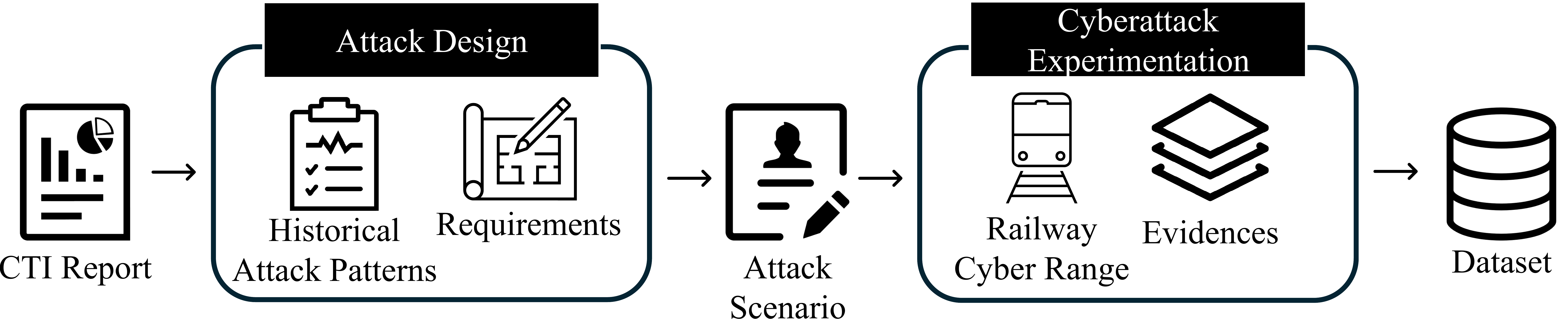}
    \caption{The methodology to construct \ac{ics} dataset}
    \label{fig:methodology}
\end{figure*}

To construct a dataset that is representative of \ac{ics} attack, we describe the process
of designing and executing the attack scenario. As reflected in Figure~\ref{fig:methodology},
this is segregated into two phases. The first phase involves the process of designing 
the cyberattack while the second phase performs experimentation which orchestrates 
the cyberattack on the railway cyber range.
To conduct attack scenario that are representative of real-world cyberattacks,
we utilize the information found from \ac{cti} reports that describe past incidents.
This information is analyzed to identify 
historical attack patterns that contribute to attack complexity and realism. This
leads to the process of collecting the relevant information related to the infrastructure, 
malware, and activity as a requirement. The outcome of the first phase produces 
the attack scenario which describes the timeline of the cyberattack.
The second phase utilizes the information in the attack scenario to execute the cyberattack
on the railway platform. Upon executing the attack scenario, the resulting evidences 
are extracted from the railway platform. These evidences are collected as a dataset that 
represents the observations of the \ac{ics} cyberattack.

\subsection{Phase I: Attack Design}
To design scenarios with sufficient complexity, it is vital to capture the attack patterns that 
are observed from existing cyberattacks. We begin by identifying the relevant information with
regards to historical attack patterns.
To attain a realistic cyberattack, we utilize the observables that are described in \ac{cti} reports.
This includes information related to the threat actors, the discovered \ac{ioc}, and the
attack activities in terms of \ac{ttp} that are observed from past cyberattacks. 
The \ac{cti} report describes the overview of the threat actor, stating about their background
and motivation behind the attack on the critical infrastructure. Additionally, \ac{cti} report
describes the \ac{ioc} that are observed from the cyberattack such as the IP addresses 
(\eg communication with \ac{c2} server) and file hashes of suspicious artifacts. The \ac{ttp}
describe the threat actor's behaviors and methods that are observed in the infrastructure. 
Due to the variety of \ac{cti} reports that describe the same cyberattack, the key challenge is to 
identify non-conflicting information that contributes towards the appearance of indicators 
in the resulting evidence.
Despite this, the information described in \ac{cti} report describes complementary aspects of 
the cyber threat, serving as the basis for designing attack scenarios. This provides a comprehensive 
understanding of historical \ac{ics} cyberattacks, especially for incidents that happen on railway systems.

To design a realizable end-to-end cyberattack, we gather the information about the required components and 
activities. One challenge is to identify the relevant details from unstructured information, which 
are then collected as a requirement. There are various types of components
that are needed to realize the cyberattack. This includes the process of defining the network segments
along with their respective hosts. Additionally, network-related systems (\eg router, firewall) can be 
specified to allow or deny access in terms of network connectivity. When the infrastructure
is set, the next step is to implement the behavior component which provides interactivity among the systems.
The attack behaviors are defined as a sequence of actions to be taken by the threat actor (\eg running commands,
executing malware). To increase the complexity of finding malicious activities, one challenge is to 
design diverse benign behaviors and include them 
as part of the requirements. These benign behaviors replicate the day-to-day activities of humans 
(\eg administrative staff, technician) that utilize the system. This enhances the realism of the cyberattack
as contextually relevant activities are included as part of the design.

Upon integrating the historical attack patterns as part of the requirements, the components are
then translated to an attack scenario. This scenario contains information that can be classified into
three categories, namely the network segments, hosts, and activities. This contains the technical description
about the components that are then used to instantiate cyberattack. For network segments, we define the 
logical groups of network segments that represent the real-world physical networks (\eg external network, 
operations network, physical network). This includes network configurations such as subnets and IP addresses. 
Additionally, the hosts are defined with configurations related to the operating system 
(\eg operating system type and version, credentials) and applications (\eg productivity software).
To simulate the activities, the benign and attack behaviors are chronologically defined as the actions taken 
by an actor at a specific time (\eg hacker runs a command to find the credential file at 08:50).

\subsection{Phase II: Cyberattack Experimentation}

To reproduce the attack, the experimentation phase uses the attack scenario to instantiate the cyberattack 
on a testbed that is connected to the railway cyber range. The testbed provides the means
to realize virtual instances in their respective network segments. This enables the simulation of 
computing infrastructure that is connected to the railway platform. Additionally, the railway platform mirrors 
the physical railway infrastructure with signaling systems and rolling stocks across different tracks.
To simulate the benign and attack activities, we utilize \ac{cue}~\cite{cuegithub} which uses customizable
scheduling profiles to generate benign and malicious traffic according to the attack scenario. Overall, 
the testbed serves as a sandbox for simulation, analysis, and generating data related to both computing 
and railway infrastructure.

The cyberattack simulation generates the corresponding data that is useful for analysis. To obtain the 
relevant evidence, we extract the data that is available on the testbed. The key challenge in this phase 
is to ensure that the resulting indicators are preserved in the extracted data. We utilize \verb|tcpdump| to 
capture the network traffic in the routers and switches. This provides an overview of internal communications 
between hosts and communications across network segments. The network traffic is captured in \verb|pcap| format, 
which is compatible with various network forensic analysis tools (\eg Wireshark). For system-level evidence, 
we collect the state of the affected hosts by taking a memory snapshot. The snapshot provides an insight into 
the attack activities that occurred in the affected hosts (\eg running processes, loaded libraries, and open sockets)
and can be analyzed using memory forensic tools (\eg Volatility). Additionally, we capture the disk images of the 
affected hosts. The disk images are captured as E01 format which is widely used in \ac{dfir} investigations due 
to its forensically sound characteristics (\ie metadata and integrity verification). The evidences are extracted 
from the testbed and collected as a dataset. The indicators that are captured during the experimentation are useful
for analysis as they represent the observations of the simulated cyberattack.

\section{Experiment}
\label{sec:experiment}

In this section, we design two attack scenarios to simulate \ac{ics} cyberattacks
that involve the railway cyber range. We then empirically evaluate 
the evidence that is extracted from the \ac{ics} infrastructure. The dataset 
includes evidence that is appropriate for \ac{dfir}
analysts to conduct forensics on filesystems, memory, network, and logs.

\subsection{Scenario For \ac{ics} Cyberattack}

To demonstrate the usefulness of the datasets, we designed two attack scenarios, 
namely \sone and \stwo, as described below. These scenarios are designed to simulate
end-to-end cyberattack throughout the \ac{ics} environment.

\subsubsection{Attack Scenario 1}

\begin{figure*}[htp!]
    \centering
	\includegraphics[width=\textwidth]{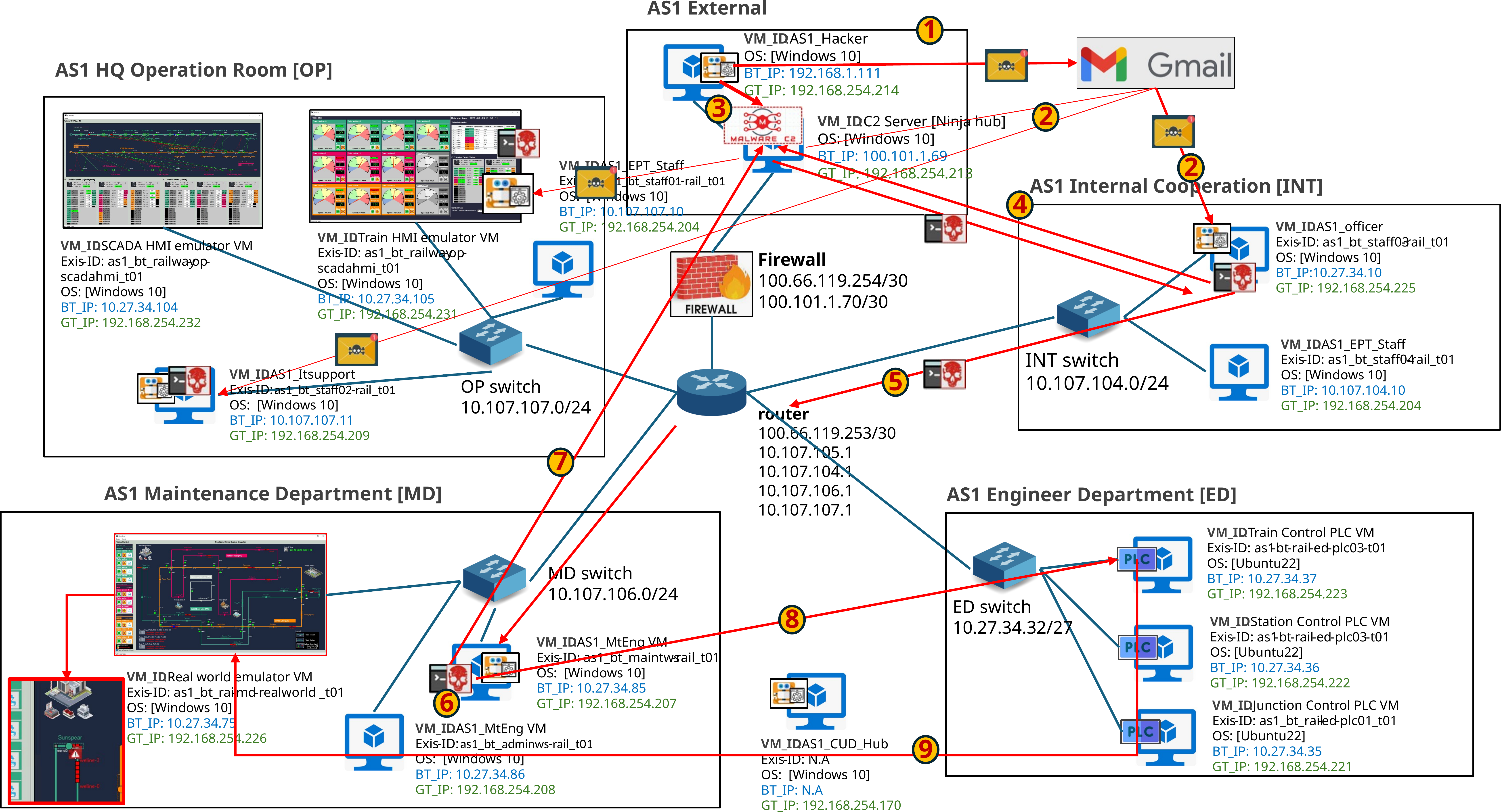}	
    \caption{Environment setup and attack path for \sone}
    \label{fig:env-s1}
\end{figure*}

-- We design the cyberattack scenario named \sone to demonstrate a multi-stage 
\ac{ics} cyberattack, specifically on a railway infrastructure as reflected in 
Figure~\ref{fig:env-s1} along with its attack steps as described in 
Table~\ref{tab:attack-s1}. We studied and implemented the attack patterns 
(\eg execution of custom \ac{ics} malware) that are commonly observed from 
\ac{ics} cyberattacks conducted by \ac{apt} groups (\eg Dragonfly and Allanite \ac{apt}s).
Based on the historical attack patterns, we integrate the patterns as part of the 
requirements for the \sone attack scenario. This scenario consists of five network segments, 
four of which form the internal \ac{ics} networks. The first segment is the external network 
which simulates two hosts from a public network. These hosts are controlled by the
threat actor that connects from the internet. The external network is connected to 
the internal \ac{ics} networks through a firewall that filters unauthorized inbound 
network traffic. The second network segment is the internal corporation network that consists of 
two hosts used by railway staff. Subsequently, the headquarters operation room network contains 
four hosts equipped with consoles for real-time control and monitoring by headquarters operators 
and engineers. This is followed by the maintenance department network which consists of 
three hosts that manage the diagnostics of the railway infrastructure and rolling stock. 
The engineer department network consists of three hosts that simulates the \ac{plc} which provides 
functionalities to the railway infrastructure (\ie train control, station control, and junction control).

\begin{table}[htp!]
\centering
\begin{tabular}{@{}cl@{}}
\toprule
\textbf{Step} & \multicolumn{1}{c}{\textbf{Description}}                                                                                                                                                                           \\ \midrule
1             & \begin{tabular}[c]{@{}l@{}}An insider downloads the spyTrojan malware into a system\\ within the internal corporation network\end{tabular}                                                                         \\
2             & \begin{tabular}[c]{@{}l@{}}The malware is disguised as \verb|updateInstaller.exe|, and is received \\ via email and subsequently propagated to other staff members \\ through internal email communication\end{tabular} \\
3             & \begin{tabular}[c]{@{}l@{}}The hacker executes malicious tasks using the spyTrojan \\ once it is installed\end{tabular}                                                                                            \\
4             & \begin{tabular}[c]{@{}l@{}}The spyTrojan exfiltrates sensitive information to the C2 server, \\ including user credentials, keystrokes, and screenshots\end{tabular}                                               \\
5             & \begin{tabular}[c]{@{}l@{}}The spyTrojan securely copies the FCI module to the PLC \\ machine using SCP\end{tabular}                                                                                               \\
6             & \begin{tabular}[c]{@{}l@{}}The spyTrojan remotely executes the FCI module on the \\ compromised machine using SSH\end{tabular}                                                                                     \\
7             & \begin{tabular}[c]{@{}l@{}}The FCI module sends the Modbus data readings from the PLC to \\ the C2 server and resolves the FCI attack targets and parameters\end{tabular}                                           \\
8             & \begin{tabular}[c]{@{}l@{}}The FCI module frequently sends false Modbus TCP coil control \\ commands to the PLC\end{tabular}                                                                                       \\
9             & \begin{tabular}[c]{@{}l@{}}The PLC disables the collision avoidance control system and \\ triggers an emergency power cutoff to the 'weline01' train\end{tabular}                                                  \\ \bottomrule
\end{tabular}%
\caption{Attack steps for \sone attack scenario}
\label{tab:attack-s1}
\end{table}

To simulate the attack activities for \sone, we construct two custom malware as part of the attack steps
as described in Table~\ref{tab:attack-s1}. The first malware is \textit{spyTrojan} which is an
espionage and delivery toolkit for the threat actor to monitor and transfer data from its victim. The 
second malware is \textit{\ac{fci} module} which monitors and executes Modbus commands base on 
instructions from the threat actor.

\subsubsection{Attack Scenario 2}
\label{sec:as2}

-- We introduce the design of the second cyberattack scenario named \stwo. To construct 
a scenario with realistic attack patterns, we refer to a specific \ac{apt} group 
called APT44, also commonly known as Sandworm. The \ac{apt} group has conducted 
their campaigns since at least 2014 and are known to attack \ac{it} and critical 
infrastructures. In this case, we refer to a specific cyberattack conducted 
in 2015 that utilized the updated BlackEnergy malware~\cite{FIROOZJAEI2022100487} 
to perform \ac{ddos} attack on a power grid, temporarily disrupting the power supply 
to the public. In \stwo, we introduce a new dimension to the \ac{ics} control systems 
by adding a power grid simulation system to the railway system.

\begin{figure}[htp!]
    \centering     
	\includegraphics[width=\textwidth]{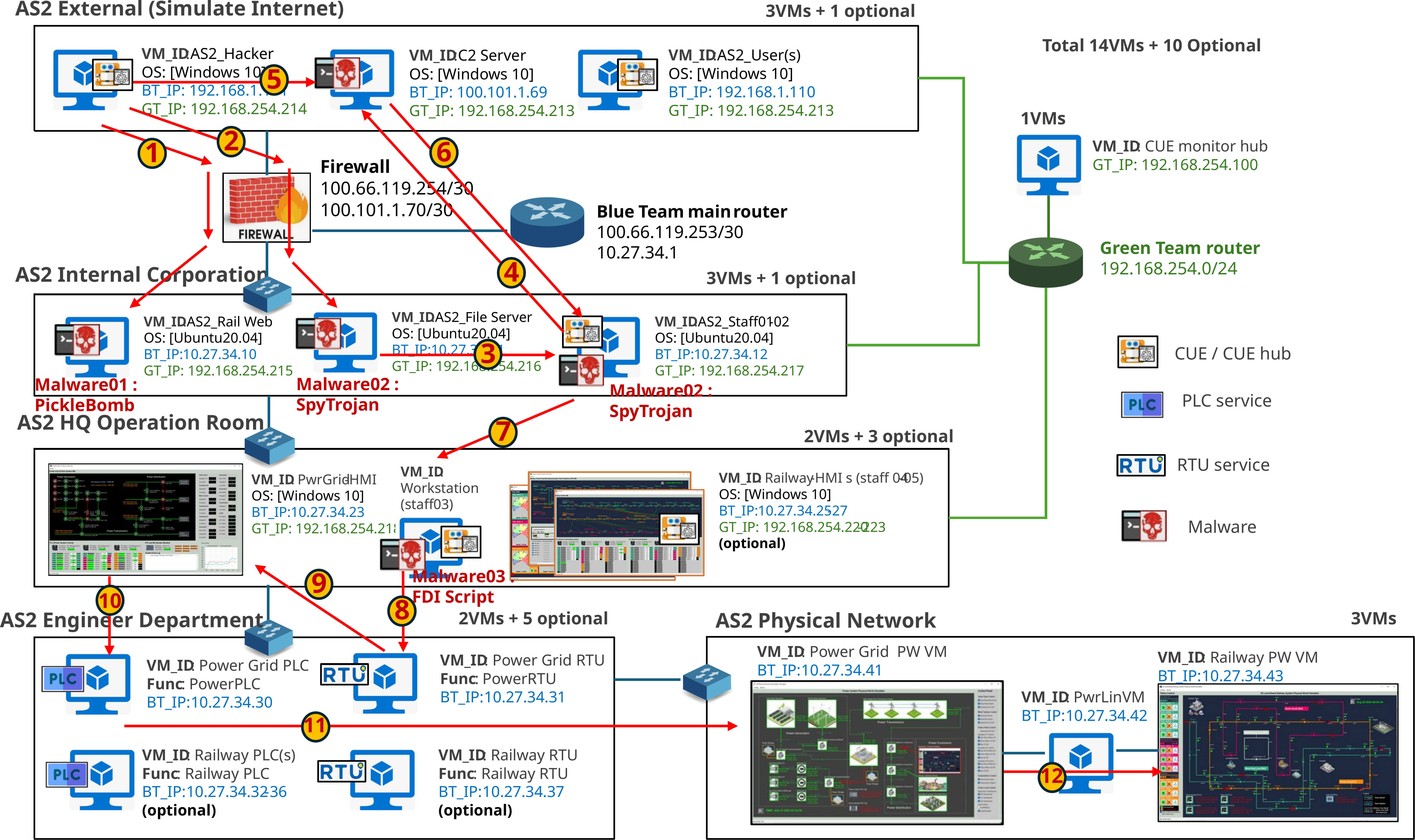}	
    \caption{Environment setup and attack path for \stwo}
    \label{fig:env-s2}
\end{figure}

The \stwo attack scenario is set up with five network segments in the environment setup as 
reflected in Figure~\ref{fig:env-s2}, along with the attack steps as described in 
Table~\ref{tab:attack-s2}. The first network segment is the external network
which includes hosts (\eg threat actor, \ac{c2} Server, users) that are from a public network.
This is followed by the internal corporation network which consists of at least three hosts
that simulate the day-to-day activities (\eg hosting of the railway web application, 
FTP file sharing, railway officer's routine). This network is isolated from the public 
and connects through a firewall that restricts unauthorized access, serving as the sole 
entry point into the critical infrastructure. This network is connected to the headquarters 
operation network which contains at least two hosts equipped with consoles for control and 
monitoring by headquarters operators and engineers. 

Subsequently, the engineer department network
consists of at least two essential hosts, which include the \ac{plc} that controls the power breakers 
for the power grid simulation and \ac{rtu} that collects power grid operational data. This network is 
directly connected to the physical network that consists of three major components, namely
the power grid simulation, the railway simulation, and the power link. The power grid simulation 
include hosts that simulate the real world functionalities such as communicating the state of the 
power breaker to the \ac{plc}, generating metering units data, and supplying power to the 
railway simulation system. The railway simulation consists of hosts that perform railway functionalities, 
including the railway power distribution system.

\begin{table}[htp!]
\begin{tabular}{@{}cl@{}}
\toprule
\textbf{Step} & \multicolumn{1}{c}{\textbf{Description}}                                                                                                                                                                                                                                                  \\ \midrule
1             & \begin{tabular}[c]{@{}l@{}}Threat actor exploits a deserialization vulnerability in the image upload feature \\ within the railway web application, allowing them to execute malicious codes, \\ access sensitive files, and obtain credentials to a FTP file sharing server\end{tabular} \\
2             & \begin{tabular}[c]{@{}l@{}}Threat actor uses a valid credential to upload a spy trojan disguised as \\ \verb|ZoomMeetingInstaller.exe| in the file sharing server.\end{tabular}                                                                                                                \\
3             & \begin{tabular}[c]{@{}l@{}}Staff synchronizes their local folder with the file sharing server, unknowingly \\ downloaded and executed the disguised spy trojan to start an online meeting.\end{tabular}                                                                                   \\
4             & \begin{tabular}[c]{@{}l@{}}The spy trojan collects information from the victim and communicates its findings\\ with the C2 server where the threat actor has control using a web interface.\end{tabular}                                                                                  \\
5             & \begin{tabular}[c]{@{}l@{}}The spy trojan is directed by the C2 server to perform network reconnaissance,\\ stealing sensitive files, and records keystrokes and screen with the goal of finding\\ an IT engineer PC to access SCADA workstation.\end{tabular}                          \\
6             & \begin{tabular}[c]{@{}l@{}}The threat actor devises a S7Comm-based FDI attack script based on stolen\\ manual, then utilize the spy trojan to download the script onto the victim to \\ effectively bypass firewall protection\end{tabular}                                               \\
7             & \begin{tabular}[c]{@{}l@{}}The threat actor uses the stolen credential to transfer the FDI attack script from \\ the IT engineer's PC to the SCADA workstation using SCP, which provides \\ remote access to the SCADA workstation.\end{tabular}                                          \\
8             & \begin{tabular}[c]{@{}l@{}}The FDI script connects to the RTU and injects false voltage and current data\\ into its memory.\end{tabular}                                                                                                                                                  \\
9             & \begin{tabular}[c]{@{}l@{}}The HMI reads the false RTU values and detected an anomaly, thus triggering the\\ power transformer's automated protection mechanism.\end{tabular}                                                                                                             \\
10            & \begin{tabular}[c]{@{}l@{}}The HMI changes to alert state, temporarily halt PLC auto control and issue the \\ command to turn off circuit breaker to protect the transformer.\end{tabular}                                                                                                \\
11            & \begin{tabular}[c]{@{}l@{}}The PLC received the Modbus command and turn off the circuit breaker in the \\ power grid physical world simulation.\end{tabular}                                                                                                                              \\
12            & This causes the railway power outage as the power value drops to zero.                                                                                                                                                                                                                    \\ \bottomrule
\end{tabular}
\caption{Attack steps for \stwo attack scenario}
\label{tab:attack-s2}
\end{table}

To execute the cyberattack for \stwo, we construct three custom malware as part of the attack steps
as described in Table~\ref{tab:attack-s2}. The first malware is \textit{pickle bomb} which is a 
serialized web shell backdoor that is delivered through an image upload vulnerability. This is used 
by the threat actor to execute malicious commands on the web server as described in the first attack step. 
The second malware is \textit{spy trojan} that receives instructions from the \ac{c2} server to perform 
network scan, lateral movement, and file transfers. The third malware is \textit{S7Comm FDI Script} which 
is used to perform false injection attack at high frequency with the purpose of overwriting the \ac{rtu} memory.

\subsection{Dataset Construction}

The evidences from the respective attack scenario are extracted from the testbed and collected as a dataset.
There are four types of evidence, namely disk image, memory image, network capture, and system-level logs. In 
addition to the captured dataset, we provide the screen recording of the respective emulators as supplementary 
information that visualizes activities occured during the cyberattack. We also include the \ac{ssl} key 
that can be used to decrypt the \ac{ssl} communication with the \ac{c2} server and categorize the 
\ac{ssl} key logs as part of the system-level logs.

\begin{table}[htp!]
\begin{tabular}{@{}cccccc@{}}
\toprule
                              &                                                                     & \textbf{File System} & \textbf{Memory Image} & \textbf{Network Capture} & \textbf{System Logs} \\ \midrule
\multirow{2}{*}{\textbf{AS1}} & \textbf{Total Size}                                                 & 19.63 GB             & 7.00 GB                  & 2.13 GB                  & \ding{55}                 \\
                              & \textbf{\begin{tabular}[c]{@{}c@{}}Indicator \\ Count\end{tabular}} & 2                    & 13                    & 14                       & \ding{55}                 \\
\multirow{2}{*}{\textbf{AS2}} & \textbf{Total Size}                                                 & \ding{55}                 & 4.64 GB               & 966.77 MB                 & 352.92 KB             \\
                              & \textbf{\begin{tabular}[c]{@{}c@{}}Indicator \\ Count\end{tabular}} & \ding{55}                 & 14                    & 8                        & 6                    \\ \bottomrule
\end{tabular}
\caption{Dataset statistics of the respective attack scenario}
\label{tab:dataset-stat}
\end{table}

The dataset statistics of the respective attack scenario are presented in Table~\ref{tab:dataset-stat}.
The statistics show the total approximate archived file size of the evidence from the respective attack scenarios. 
Additionally, this includes the minimum amount of indicators that are found from the respective evidence.  

The \sone dataset consists of one disk image, two memory images, and one network dump that is collected across 
two days. The disk image is taken 
from the maintenance workstation (\ie \verb|10.27.34.85|) in the maintenance department network immediately 
after the \verb|welin01| train has crashed at the end of the attack steps. To capture the state of the affected 
systems, we take a memory dump from two hosts after the train has crashed, namely the same maintenance workstation 
and the \verb|staff03| workstation (\ie \verb|10.27.34.10|) in the internal corporation network. In addition to 
system-level data, we capture the network traffic from all interfaces firewall and router between 03 April 2024 and 
04 April 2024. The captured network traffic consists of all communications between hosts in the internal network 
and with hosts in the external network. To supplement the dataset, we provide a video that reflects the 
cyberattack activities that occurred in the \verb|Real World| emulator from the maintenance department network.
The system-level logs for \sone are unavailable.

The \stwo dataset consists of two memory images, one network capture, and one application log. The memory dump is 
taken from two hosts, namely the \verb|staff01| and \verb|staff03| workstations in the internal corporation network. 
The network capture consists of network traffic monitored from the main router (\ie\verb|100.66.119.253/30|). 
Additionally, we capture the application logs for the railway company's web application. We also provide three videos 
that show the cyberattack activities that occurred in the \verb|Power Grid HMI| from the internal corporation network 
and the \verb|Power Grid| as well as \verb|Physical World| from the physical world network.
The disk image for \stwo is unavailable.

\subsection{Case Study: \stwo Attack Scenario}

This section conducts a case study of \stwo to showcase the usefulness of the dataset for \ac{ics} cyberattack analysis.
Based on the attack scenario as described in Section~\ref{sec:as2}, we analyze the evidence that is extracted from 
affected hosts and identify the \ac{ioc}. As described in the attack steps (\ie Table~\ref{tab:attack-s2}), 
the threat actor interacts with 
a public-facing web application and perform a file upload during the early attack stages. The file upload activity 
is captured in the web application logs, as reflected in Listing~\ref{lst:case-weblog}. While the filename 
\verb|image.txt| appears to be benign, the contents of the uploaded file indicate a resemblance to an encoded 
text. When the content is decoded using \verb|base64| format, it becomes apparent that it contains malicious code 
to create a Flask-based web shell written in Python.

\begin{lstinputlisting}[caption={Logs from web application showing file upload}, 
                        label={lst:case-weblog},
                        basicstyle=\scriptsize\ttfamily,
                        captionpos=b,
                        frame=single
                        ]{figures/case-weblog.out}
\end{lstinputlisting}

In subsequent attack steps, the trojan (\ie \verb|ZoomMeetingInstaller.exe|) is executed on 
the first victim's host. Since the trojan is 
executed as a process in the memory, the memory image taken from the host of the affected victim captures this 
\ac{ioc}. As reflected in Listing~\ref{lst:case-memory}, executing \verb|ZoomMeetingInstaller.exe| results 
in the creation of process 8340 on the system. This reflects the successful execution of the trojan, allowing 
the victim to communicate with the \ac{c2} server.

\begin{lstinputlisting}[caption={Running processes based on the memory image taken from the first victim}, 
                        label={lst:case-memory},
                        basicstyle=\scriptsize\ttfamily,
                        captionpos=b,
                        frame=single
                        ]{figures/case-memory.out}
\end{lstinputlisting}

\begin{figure}[htp!]
    \centering
    \setlength{\FrameSep}{0pt}
    \begin{framed}
        \includegraphics[width=\textwidth]{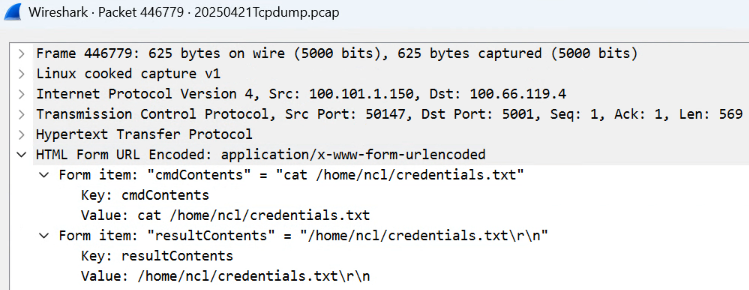}
    \end{framed}
    \caption{Communication between the victim and the \ac{c2} server}
    \label{fig:case-network}
\end{figure}

When the installed malware communicates with the \ac{c2} server, the captured network traffic contains the 
network-level \ac{ioc} that reflects their communication. Based on Figure~\ref{fig:case-network}, the network packet 
indicates that the communication contains a suspicious encoded URL string. When the string is decoded, it 
indicates the intention to exfiltrate the content of \verb|credentials.txt| by using the \verb|cat| command. 
This reflects the attack steps by collecting sensitive information from the victim and send them to the threat actor.

We have also included three videos as part of \stwo dataset, showing the failure of the power grid which affects 
the entire railway infrastructure. This marks the success for the threat actors in reaching their desired goal 
by causing a disruption to the railway infrastructure according to \stwo attack steps.
Based on the forensics performed on the evidences, analysts are able to identify various \ac{ioc} from 
system-level and network-level perspectives. This allows analysts to reconstruct the attack sequence based 
on the discovered indicators, effectively understand the threat actor's behaviors and the motivation behind 
the cyberattack.

\section{Related Work}
\label{sec:related-work}

We present two topics that are related to this work, namely existing works that construct a virtual model to simulate 
\ac{ics} infrastructure and works that involves the generation and sharing of \ac{ics}-related datasets.

\subsection{Virtual Modeling of \ac{ics} Infrastructure}
A virtual model of \ac{ics} infrastructure serves as a platform for safe experimentation of cyber threats, thus providing a means 
to obtain insights into 
\ac{ics} cyberattacks. An open-source framework called SCASS~\cite{10.1016/j.cose.2025.104315} provide a means for analysts 
to construct a customizable and extensible testbed that replicates complex \ac{scada} and \ac{ics} infrastructure with high fidelity.
Similarly, ICSSIM~\cite{DEHLAGHIGHADIM2023103906} proposes a framework to construct a virtual \ac{ics} testbed, and provides a means 
to deploy them on actual hardware (\eg physical devices, containerized environments, simulation systems). With a focus on generating 
data that is appropriate for analysis, an existing work~\cite{10718749} proposes a testbed for repeatable cyberattack simulations 
aimed at generating useful datasets. Likewise, the Tennessee-Eastman System simulates a chemical process that is modeled
after a real industrial operational systems~\cite{DOWNS1993245}, thus enabling the generation of data from attack simulation 
(\eg PLC-SAGE~\cite{10495752}). While existing works provide a general platform to simulate \ac{ics} cyberattacks, the 
SAFETY4RAILS project~\cite{safety4rails} and CaESAR~\cite{10.1007/978-3-031-25460-4_17} focus on providing a simulation platform 
for the railway infrastructure. In this work, we propose a railway cyber range that models a railway infrastructure, thus 
providing a platform to simulate complex \ac{ics} cyberattacks, effectively producing indicators that can be extracted as datasets.

\subsection{\ac{ics} Cyberattack Dataset}
High-quality datasets are essential to facilitate research and analysis of cyberattacks on \ac{ics}. To address this, 
the HAI dataset 1.0~\cite{256936} provide a \ac{cps} dataset based on numerous benign and attack scenarios. 
Similarly, the SWaT~\cite{10.1007/978-3-319-71368-7_8}, WaDi~\cite{10.1145/3055366.3055375}, and EPIC~\cite{10.1007/978-3-030-69781-5_9}
datasets are constructed based on \ac{ics}-based attack scenarios executed on water treatment, water distribution, and 
energy control respectively. Additionally, 
ICS-ADD~\cite{10516443} provides an open-source dataset that contains network traffic captured from \ac{ics} systems that 
are subjected to simulated cyberattacks. Instead of generating datasets from cyberattack simulations, 
this work~\cite{electronics13101920} proposes a data augmentation approach that leverages existing data to recreate 
a simulation environment, thus providing a means to produce a realistic synthetic dataset. In this work, we provide 
railway-specific \ac{ics} datasets that are generated based on cyberattack simulation conducted on a railway cyber range.
\section{Conclusion}
\label{sec:conclusion}

To enable cyberattack analysis of trending \ac{ics} cyberattacks, we construct railway-specific 
\ac{ics} datasets that are generated from our railway cyber range. Specifically, we design cyberattack scenarios 
based on commonly observed 
attack patterns from past incidents. The cyberattack scenarios are realized on the railway cyber range to 
simulate the \ac{ics} cyberattack. The data and indicators that result from the cyberattack simulation are extracted 
as evidence and collected as a dataset. The \ac{ics} dataset is publicly shared to facilitate research and analysis 
in \ac{ics} cyberattacks.

\bibliographystyle{splncs04}

\end{document}